\documentclass{ws-rv9x6}
\usepackage{ws-rv-van}             
\makeindex
\begin{document}

\chapter{Omega and the time evolution of the $n$-body problem\label{chapter-svozil}}

\author[K. Svozil]{Karl Svozil}
\address{Institut f\"ur Theoretische Physik, University of Technology Vienna,\\
Wiedner Hauptstra\ss e 8-10/136, A-1040 Vienna, Austria\\
svozil@tuwien.ac.at}

\begin{abstract}
The series solution of the behavior of a finite number of physical bodies and Chaitin's Omega number
share quasi-algorithmic expressions; yet both lack a computable radius of convergence.
\end{abstract}


\body

\section{Solutions to the $n$--body problem}

The behaviour and evolution of a finite number of bodies is a sort of ``rosetta stone''
of classical celestial mechanics insofar as its investigation induced a lot of
twists, revelations and unexpected issues.
Arguably the most radical deterministic position on the subject
was formulated by Laplace,
stating that \cite[Chapter II]{laplace-prob}
{\em
``We ought then to regard the present state of the
universe as the effect of its anterior state and as the
cause of the one which is to follow. Given for one
instant an intelligence which could comprehend all the
forces by which nature is animated and the respective
situation of the beings who compose it an intelligence
sufficiently vast to submit these data to analysis it
would embrace in the same formula the movements of
the greatest bodies of the universe and those of the
lightest atom; for it, nothing would be uncertain and
the future, as the past, would be present to its eyes.''
}

In what may be considered as the beginning of deterministic chaos theory,
Poincar{\'e} was forced to accept a gradual departure from the deterministic position:
sometimes small
variations in the initial state of the bodies could lead to huge variations in their
evolution in later times.
In Poincar{\'e}'s own words  \cite[Chapter 4, Sect. II, pp. 56-57]{poincare14},
{\em
``If we would know the laws of Nature and the state of the Universe precisely
for a certain time,
we would be able to predict with certainty
the state of the Universe for any later time.
But
[[~$\ldots$~]]
it can be the case that small differences in the initial values
produce great differences in the later phenomena;
a small error in the former may result in a large error in the latter.
The prediction becomes impossible and we have a `random phenomenon.'
''}

In what follows we present an even more radical departure from Laplacian determinism.
A physical system of a finite number of bodies capable of universal computation will be presented
which has the property that certain propositions remain not only provable
intractable, but provable unknowable.
Pointedly stated, our knowledge of any such system remains incomplete forever.
For the sake of making things worse, we shall ``compress'' and ``compactify''
this kind of physical incompleteness by considering physical observables
which are truly random, i.e., algorithmically incompressible and stochastic.

The methods of construction of
physical $n$--body observables exhibiting the above features turn out to be rather humble and straightforward.
In a first step, it suffices to reduce the problem to the halting problem for universal computation.
This can be achieved by ``embedding'' a universal computer into a
suitable physical system of a finite number of bodies.
The associated ballistic computation will be presented in the next section.
In a second reduction step, the universal computer will be directed to attempt to ``compute''
Chaitin's Omega number, which is provable random,
and which is among the ``most difficult'' tasks imaginable.
Finally, consequences for the series solutions
 \cite{Sundman12,Wang91,Diacu96,Wang01} to the general $n$-body problem will be discussed.

\section{Reduction by ballistic computation}

In order to embed reversible universal computation into a quasi-physical environment,
Fredkin and Toffoli introduced a ``billiard ball model''
\cite{fred-tof-82,margolus-billard,margolus-02,Adamatzky-02}
based on the collisions of spheres as well as on mirrors reflecting the spheres.
Thus collisions and reflections are the basic ingredients for building universal computation.

\begin{figure*}
\begin{center}
\begin{tabular}{ccccccc}
\unitlength .4mm 
\linethickness{0.4pt}
\ifx\plotpoint\undefined\newsavebox{\plotpoint}\fi 
\begin{picture}(115,130)(0,0)
\qbezier(10,10)(30.5,80.5)(60,80)
\qbezier(10,110)(30.5,39.5)(60,40)
\qbezier(109.5,10)(89,80.5)(59.5,80)
\qbezier(109.5,110)(89,39.5)(59.5,40)
\put(35.5,98.25){\vector(1,3){.18}}\multiput(34.07,94.32)(.0833,.25){3}{\line(0,1){.25}}
\multiput(34.57,95.82)(.0833,.25){3}{\line(0,1){.25}}
\multiput(35.07,97.32)(.0833,.25){3}{\line(0,1){.25}}
\put(12.25,17.5){\vector(1,4){.18}}\multiput(11.07,13.57)(.0667,.25){3}{\line(0,1){.25}}
\multiput(11.47,15.07)(.0667,.25){3}{\line(0,1){.25}}
\multiput(11.87,16.57)(.0667,.25){3}{\line(0,1){.25}}
\put(12.25,102.5){\vector(1,-4){.18}}\multiput(11.07,106.07)(.0667,-.25){3}{\line(0,-1){.25}}
\multiput(11.47,104.57)(.0667,-.25){3}{\line(0,-1){.25}}
\multiput(11.87,103.07)(.0667,-.25){3}{\line(0,-1){.25}}

\put(34.25,25.25){\vector(1,-4){.18}}\multiput(33.82,26.82)(.0625,-.4375){2}{\line(0,-1){.4375}}
\multiput(9.82,9.82)(.0679,.2381){4}{\line(0,1){.2381}}
\multiput(10.37,11.73)(.0679,.2381){4}{\line(0,1){.2381}}
\multiput(10.91,13.63)(.0679,.2381){4}{\line(0,1){.2381}}
\multiput(11.45,15.54)(.0679,.2381){4}{\line(0,1){.2381}}
\multiput(12,17.44)(.0679,.2381){4}{\line(0,1){.2381}}
\multiput(12.54,19.35)(.0679,.2381){4}{\line(0,1){.2381}}
\multiput(13.08,21.25)(.0679,.2381){4}{\line(0,1){.2381}}
\multiput(13.62,23.16)(.0679,.2381){4}{\line(0,1){.2381}}
\multiput(14.17,25.06)(.0679,.2381){4}{\line(0,1){.2381}}
\multiput(14.71,26.97)(.0679,.2381){4}{\line(0,1){.2381}}
\multiput(15.25,28.87)(.0679,.2381){4}{\line(0,1){.2381}}
\multiput(15.8,30.78)(.0679,.2381){4}{\line(0,1){.2381}}
\multiput(16.34,32.68)(.0679,.2381){4}{\line(0,1){.2381}}
\multiput(16.88,34.59)(.0679,.2381){4}{\line(0,1){.2381}}
\multiput(17.42,36.49)(.0679,.2381){4}{\line(0,1){.2381}}
\multiput(17.97,38.4)(.0679,.2381){4}{\line(0,1){.2381}}
\multiput(18.51,40.3)(.0679,.2381){4}{\line(0,1){.2381}}
\multiput(19.05,42.21)(.0679,.2381){4}{\line(0,1){.2381}}
\multiput(19.6,44.11)(.0679,.2381){4}{\line(0,1){.2381}}
\multiput(20.14,46.01)(.0679,.2381){4}{\line(0,1){.2381}}
\multiput(20.68,47.92)(.0679,.2381){4}{\line(0,1){.2381}}
\multiput(21.22,49.82)(.0679,.2381){4}{\line(0,1){.2381}}
\multiput(21.77,51.73)(.0679,.2381){4}{\line(0,1){.2381}}
\multiput(22.31,53.63)(.0679,.2381){4}{\line(0,1){.2381}}
\multiput(22.85,55.54)(.0679,.2381){4}{\line(0,1){.2381}}
\multiput(23.4,57.44)(.0679,.2381){4}{\line(0,1){.2381}}
\multiput(23.94,59.35)(.0679,.2381){4}{\line(0,1){.2381}}
\multiput(24.48,61.25)(.0679,.2381){4}{\line(0,1){.2381}}
\multiput(25.02,63.16)(.0679,.2381){4}{\line(0,1){.2381}}
\multiput(25.57,65.06)(.0679,.2381){4}{\line(0,1){.2381}}
\multiput(26.11,66.97)(.0679,.2381){4}{\line(0,1){.2381}}
\multiput(26.65,68.87)(.0679,.2381){4}{\line(0,1){.2381}}
\multiput(27.2,70.78)(.0679,.2381){4}{\line(0,1){.2381}}
\multiput(27.74,72.68)(.0679,.2381){4}{\line(0,1){.2381}}
\multiput(28.28,74.59)(.0679,.2381){4}{\line(0,1){.2381}}
\multiput(28.82,76.49)(.0679,.2381){4}{\line(0,1){.2381}}
\multiput(29.37,78.4)(.0679,.2381){4}{\line(0,1){.2381}}
\multiput(29.91,80.3)(.0679,.2381){4}{\line(0,1){.2381}}
\multiput(30.45,82.21)(.0679,.2381){4}{\line(0,1){.2381}}
\multiput(31,84.11)(.0679,.2381){4}{\line(0,1){.2381}}
\multiput(31.54,86.01)(.0679,.2381){4}{\line(0,1){.2381}}
\multiput(32.08,87.92)(.0679,.2381){4}{\line(0,1){.2381}}
\multiput(32.62,89.82)(.0679,.2381){4}{\line(0,1){.2381}}
\multiput(33.17,91.73)(.0679,.2381){4}{\line(0,1){.2381}}
\multiput(33.71,93.63)(.0679,.2381){4}{\line(0,1){.2381}}
\multiput(34.25,95.54)(.0679,.2381){4}{\line(0,1){.2381}}
\multiput(34.8,97.44)(.0679,.2381){4}{\line(0,1){.2381}}
\multiput(35.34,99.35)(.0679,.2381){4}{\line(0,1){.2381}}
\multiput(35.88,101.25)(.0679,.2381){4}{\line(0,1){.2381}}
\multiput(36.42,103.16)(.0679,.2381){4}{\line(0,1){.2381}}
\multiput(36.97,105.06)(.0679,.2381){4}{\line(0,1){.2381}}
\multiput(37.51,106.97)(.0679,.2381){4}{\line(0,1){.2381}}
\multiput(38.05,108.87)(.0679,.2381){4}{\line(0,1){.2381}}
\multiput(9.82,109.82)(.0679,-.2381){4}{\line(0,-1){.2381}}
\multiput(10.37,107.92)(.0679,-.2381){4}{\line(0,-1){.2381}}
\multiput(10.91,106.01)(.0679,-.2381){4}{\line(0,-1){.2381}}
\multiput(11.45,104.11)(.0679,-.2381){4}{\line(0,-1){.2381}}
\multiput(12,102.21)(.0679,-.2381){4}{\line(0,-1){.2381}}
\multiput(12.54,100.3)(.0679,-.2381){4}{\line(0,-1){.2381}}
\multiput(13.08,98.4)(.0679,-.2381){4}{\line(0,-1){.2381}}
\multiput(13.62,96.49)(.0679,-.2381){4}{\line(0,-1){.2381}}
\multiput(14.17,94.59)(.0679,-.2381){4}{\line(0,-1){.2381}}
\multiput(14.71,92.68)(.0679,-.2381){4}{\line(0,-1){.2381}}
\multiput(15.25,90.78)(.0679,-.2381){4}{\line(0,-1){.2381}}
\multiput(15.8,88.87)(.0679,-.2381){4}{\line(0,-1){.2381}}
\multiput(16.34,86.97)(.0679,-.2381){4}{\line(0,-1){.2381}}
\multiput(16.88,85.06)(.0679,-.2381){4}{\line(0,-1){.2381}}
\multiput(17.42,83.16)(.0679,-.2381){4}{\line(0,-1){.2381}}
\multiput(17.97,81.25)(.0679,-.2381){4}{\line(0,-1){.2381}}
\multiput(18.51,79.35)(.0679,-.2381){4}{\line(0,-1){.2381}}
\multiput(19.05,77.44)(.0679,-.2381){4}{\line(0,-1){.2381}}
\multiput(19.6,75.54)(.0679,-.2381){4}{\line(0,-1){.2381}}
\multiput(20.14,73.63)(.0679,-.2381){4}{\line(0,-1){.2381}}
\multiput(20.68,71.73)(.0679,-.2381){4}{\line(0,-1){.2381}}
\multiput(21.22,69.82)(.0679,-.2381){4}{\line(0,-1){.2381}}
\multiput(21.77,67.92)(.0679,-.2381){4}{\line(0,-1){.2381}}
\multiput(22.31,66.01)(.0679,-.2381){4}{\line(0,-1){.2381}}
\multiput(22.85,64.11)(.0679,-.2381){4}{\line(0,-1){.2381}}
\multiput(23.4,62.21)(.0679,-.2381){4}{\line(0,-1){.2381}}
\multiput(23.94,60.3)(.0679,-.2381){4}{\line(0,-1){.2381}}
\multiput(24.48,58.4)(.0679,-.2381){4}{\line(0,-1){.2381}}
\multiput(25.02,56.49)(.0679,-.2381){4}{\line(0,-1){.2381}}
\multiput(25.57,54.59)(.0679,-.2381){4}{\line(0,-1){.2381}}
\multiput(26.11,52.68)(.0679,-.2381){4}{\line(0,-1){.2381}}
\multiput(26.65,50.78)(.0679,-.2381){4}{\line(0,-1){.2381}}
\multiput(27.2,48.87)(.0679,-.2381){4}{\line(0,-1){.2381}}
\multiput(27.74,46.97)(.0679,-.2381){4}{\line(0,-1){.2381}}
\multiput(28.28,45.06)(.0679,-.2381){4}{\line(0,-1){.2381}}
\multiput(28.82,43.16)(.0679,-.2381){4}{\line(0,-1){.2381}}
\multiput(29.37,41.25)(.0679,-.2381){4}{\line(0,-1){.2381}}
\multiput(29.91,39.35)(.0679,-.2381){4}{\line(0,-1){.2381}}
\multiput(30.45,37.44)(.0679,-.2381){4}{\line(0,-1){.2381}}
\multiput(31,35.54)(.0679,-.2381){4}{\line(0,-1){.2381}}
\multiput(31.54,33.63)(.0679,-.2381){4}{\line(0,-1){.2381}}
\multiput(32.08,31.73)(.0679,-.2381){4}{\line(0,-1){.2381}}
\multiput(32.62,29.82)(.0679,-.2381){4}{\line(0,-1){.2381}}
\multiput(33.17,27.92)(.0679,-.2381){4}{\line(0,-1){.2381}}
\multiput(33.71,26.01)(.0679,-.2381){4}{\line(0,-1){.2381}}
\multiput(34.25,24.11)(.0679,-.2381){4}{\line(0,-1){.2381}}
\multiput(34.8,22.21)(.0679,-.2381){4}{\line(0,-1){.2381}}
\multiput(35.34,20.3)(.0679,-.2381){4}{\line(0,-1){.2381}}
\multiput(35.88,18.4)(.0679,-.2381){4}{\line(0,-1){.2381}}
\multiput(36.42,16.49)(.0679,-.2381){4}{\line(0,-1){.2381}}
\multiput(36.97,14.59)(.0679,-.2381){4}{\line(0,-1){.2381}}
\multiput(37.51,12.68)(.0679,-.2381){4}{\line(0,-1){.2381}}
\multiput(38.05,10.78)(.0679,-.2381){4}{\line(0,-1){.2381}}
\put(0,55){\makebox(0,0)[cc]{$t_i$}}
\put(120,55){\makebox(0,0)[cc]{$t_f$}}
\put(0,109){\makebox(0,0)[cc]{${\bf A}$?}}
\put(0,10){\makebox(0,0)[cc]{${\bf B}$?}}
\put(120,110){\makebox(0,0)[cc]{${\bf AB}$}}
\put(120,10){\makebox(0,0)[cc]{${\bf AB}$}}
\put(50,110){\makebox(0,0)[cc]{$\overline{{\bf A}}{\bf B}$}}
\put(50,10){\makebox(0,0)[cc]{${\bf A}\overline{{\bf B}}$}}
\put(107.25,103.25){\vector(1,4){.18}}\multiput(106.75,101.5)(.08333,.29167){6}{\line(0,1){.29167}}
\put(106.25,20){\vector(1,-4){.18}}\multiput(105.75,22)(.08333,-.33333){6}{\line(0,-1){.33333}}
\end{picture}
&$\qquad $&
\unitlength .4mm 
\linethickness{0.4pt}
\ifx\plotpoint\undefined\newsavebox{\plotpoint}\fi 
\begin{picture}(115,80.5)(0,0)
\qbezier(10,10)(30.5,80.5)(60,80)
\qbezier(109.5,10)(89,80.5)(59.5,80)
\put(12.25,17.5){\vector(1,4){.18}}\multiput(11.07,13.57)(.0667,.25){3}{\line(0,1){.25}}
\multiput(11.47,15.07)(.0667,.25){3}{\line(0,1){.25}}
\multiput(11.87,16.57)(.0667,.25){3}{\line(0,1){.25}}

\put(0,10){\makebox(0,0)[cc]{${\bf A}$}}
\put(120,10){\makebox(0,0)[cc]{${\bf A}$}}
\put(106.25,20){\vector(1,-4){.18}}\multiput(105.75,22)(.08333,-.33333){6}{\line(0,-1){.33333}}
\put(60,40){\circle*{10}}
\end{picture}
\\
a)&&b)
\end{tabular}
\end{center}
\caption{Elements of universal ballistic computation realized by
attractive $1/r$ potentials.
a) Fredkin's gate can perform logical reversibility:
bodies will appear on the right outgoing paths
if and only if bodies came in at both ${\bf A}$ and ${\bf B}$;
b) Reflective ``mirror'' element realized by a quasi-steady mass.
\label{2007-chaitin-f1}}
\end{figure*}
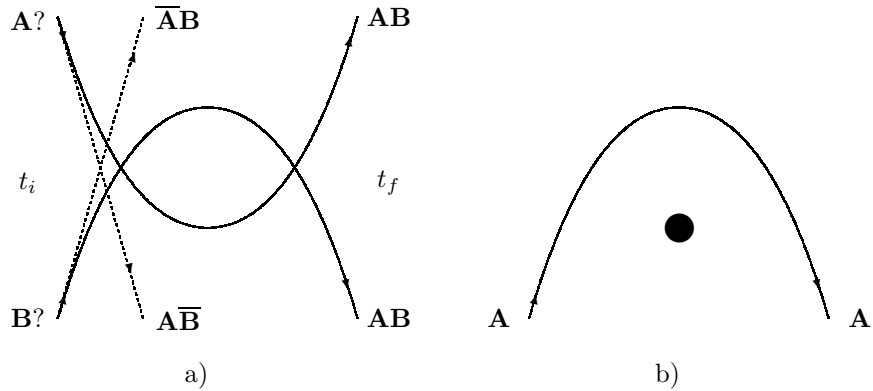

If we restrict ourselves to classical gravitational potentials without collisions,
we do not have any repulsive
interaction at our disposal; only attractive $1/r$ potentials.
Thus the kinematics corresponding to reflections and collisions has to be realized
by purely attractive interactions.
Fig.~\ref{2007-chaitin-f1}a) depicts a Fredkin gate realized by attractive interaction
which corresponds to the analogue billiard ball configuration achieved by collisions
(e.g., \cite[Fig.~4.5]{margolus-billard}).
At points
${\bf A}$
and
${\bf B}$
and time $t_i$,
two bodies are either put at both locations ${\bf A}$
and
${\bf B}$; or alternatively, one body is put
at only one location, or no bodies are placed at all.
If bodies are present at both  ${\bf A}$
and
${\bf B}$, then they will follow the right paths at later times $t_f$.
In case only one body is present at ${\bf A}$
or
${\bf B}$, only one of the dotted inner outgoing paths will be used.
Boolean logic can be implemented by the presence or absence of balls.
Fig.~\ref{2007-chaitin-f1}b) depicts a reflective ``mirror'' element realized by a quasi-steady mass.
For a proof of universality, we refer to the classical papers
on the billiard ball model cited above.

\section{Undecidability and Omega in the $n$-body problem}

By reduction to the recursive unsolvability
of the rule inference \cite{go-67,blum75blum,angluin:83,ad-91,li:92}
and the halting \cite{rogers1,odi:89,odi:99} problems,
the general induction and forecasting problem of the
$n$-body ballistic universal computer sketched above is provable unsolvable.
That is, there exist initial configurations for which it
is impossible to predict with certainty whether or not certain
``final'' states will eventually be reached.
Moreover, given a finite segment of the time evolution alone
is in general insufficient for a derivation of the initial state configuration
of the $n$-body problem.

For the sake of making things worse, we imagine an $n$-body system
attempting to evaluate its associated halting probability Omega
\cite{chaitin3,calude:94,calude-dinneen06}.
In order to establish the equivalent of prefix-free programs,
only a limited number of $n$-body initial configurations contribute to
the configuration.
Furthermore, as the computation is reversible and procedural,
certain ``final'' configurations must be defined as halting states.
This is a feature shared with the billiard ball model,
as well as with quantum computation.

\section{Consequences for series solutions}

Wang's power series solution to the $n$-body problem \cite{Wang91,Wang01}
may converge ``very slowly'' \cite{Diacu96}.
Indeed, by considering the halting problems above, and
in particular by reduction to the computation of the halting probability Omega,
certain physical observables associated with the $n$-body problem
do not have a power series solution with a {\em computable radius of convergence}.

This is a particular case of Specker's theorems in recursive analysis, stating that
there exist recursive monotone bounded sequences of rational numbers
whose limit is no computable number
\cite{Specker49}; and
there exist a recursive real function which has its maximum in the unit interval
at no recursive real number \cite{Specker57}.

It is important to realize that,
while it may be possible to evaluate
the state of the $n$ bodies by Wang's power series solution
for any finite time with a computable,
though excessively large, radius of convergence,
global observables, referring to all times, may be uncomputable.
Examples of global observables are, for instance, associated
with the stability of the solar system and associated with it,
bounds for the orbits.

This, of course,
stems from the metaphor and robustness of universal computation
and the capacity of the $n$-body problem to implement universality.
It is no particularity and peculiarity of Wang's power series solution.
Indeed, the troubles reside in the capabilities to implement Peano arithmetic and
universal computation by $n$-body problems.
Because of this capacity, there cannot exist other formalizable methods,
analytic solutions or approximations capable to decide and compute certain decision problems
or observables for the $n$-body problem.

Chaitin's Omega number,
the halting probability for universal computers,
has been invented in a totally different, unrelated algorithmic context,
and with intentions in mind which are seemingly different from issues in classical mechanics.
Thus it is fascinating that Omega is also relevant for the prediction of the behaviour
and the movement of celestial bodies.


\end{document}